\documentclass[twocolumn]{aastex63}

\usepackage{graphicx}	
\usepackage{color}
\usepackage{amsmath}	
\usepackage{amssymb}	
\usepackage{mathrsfs}
\usepackage{mathrsfs}
\usepackage{multirow}

\newcommand{\msun}{M_\odot}

\newcommand{\pc}{{\rm pc}}

\newcommand{\K}{{\rm K}}
\newcommand{\beq}{\begin{equation}}
\newcommand{\eeq}{\end{equation}}

\newcommand{\D}{{\rm d}}

\shorttitle{}
\shortauthors{}

\begin{document}

\title{Little Red Dots as the Very First Activity of Black Hole Growth}

\correspondingauthor{Kohei Inayoshi}
\email{inayoshi@pku.edu.cn}

\author[0000-0001-9840-4959]{Kohei Inayoshi}
\affiliation{Kavli Institute for Astronomy and Astrophysics, Peking University, Beijing 100871, China}

\begin{abstract}
The James Webb Space Telescope has detected massive black holes (BHs) with masses of $\sim 10^{6-8}~\msun$ within the first billion years of the universe.
One of the remarkable findings is the identification of ``Little Red Dots" (LRDs), a unique class of active galactic nuclei (AGNs) 
with distinct characteristics representing a key phase in the formation and growth of early BHs.
Here, we analyze the occurrence rate of LRDs, which emerge around redshifts $z \sim 6-8$ and sharply decline at $z < 4$. 
We find that this trend follows a log-normal distribution, commonly observed in phenomena driven by stochastic and random factors.
We propose a hypothesis that the first one or two AGN events associated with newly-formed seed BHs are observed as LRDs and 
their unique features fade in the subsequent episodes.
This naturally explains the cosmic evolution of AGN abundance over $0 < z < 5$, which follows $\propto (1+z)^{-5/2}$ due to the cumulative effect of recurring AGN activity.
The unique characteristics of LRDs are likely linked to the dense gas environments around the seed BHs, which create strong absorption features 
in the broad-line emission and enable super-Eddington accretion bursts, ultimately yielding the observed overmassive nature of BHs compared to the local relationship.
An analytical expression for the redshift evolution of LRD abundance is provided for direct comparison with future observational constraints.
\end{abstract}

\section{Introduction}

The James Webb Space Telescope (JWST) has revolutionized extragalactic research, uncovering low-luminosity AGNs at high redshifts of $z>4-7$ 
powered by accreting BHs with masses of $\sim 10^{6-8}~\msun$ \citep[e.g.,][]{Onoue_2023,Kocevski_2023,Harikane_2023_agn,Maiolino_2024_JADES, Taylor_2025a}. 
Among these ground-breaking discoveries, the identification of “Little Red Dots” (LRDs) stand out as particularly remarkable. 
LRDs are extremely compact objects ($<100~\pc$) characterized by broad emission lines on red continuum spectra, 
indicating the presence of dust-obscured AGNs \citep[e.g.,][]{Matthee_2024,Greene_2024,Labbe_2025}. 
Their cosmic abundance is several orders of magnitude higher than that of bright quasars, allowing them to be detectable 
even within JWST's narrow field of view \citep{Kokorev_2024a,Akins_2024,Kocevski_2025}. 
Under standard assumptions in measurement, these objects show BH-to-galaxy mass ratios far above the empirical values observed in the nearby universe 
\citep[e.g.,][]{Kormendy_Ho_2013,Reines_Volonteri_2015},  
suggesting that they preserve crucial information on the formation of seed BHs and their early rapid growth phases.

Despite their significance, key questions about the spectral nature of LRDs arise from multi-wavelength observations
\citep[e.g.,][]{Greene_2024,Juodzbalis_2024,Perez-Gonzalez_2024,Maiolino_2025,Wang_2024b,Wang_2025,Kocevski_2025}: 
(1) the prominent absorption feature in broad Balmer emission lines, (2) the deficit of hot dust emission, (3) the ``v-shape" SED in the rest-frame 
UV-optical range, and (4) the absence of X-ray detections. 
These characteristics indicate that LRDs might represent a unique phase of AGN activity in their formation, and 
a potential contribution from super-Eddington accretion \citep[][references therein]{Inayoshi_ARAA_2020,Volonteri_2021}.

Throughout this paper, we assume a flat $\Lambda$ cold dark matter (CDM) cosmology consistent with the constraints from Planck \citep{Planck_2016};
$h = 0.677$, $\Omega_{\rm m}= 0.307$, $\Omega_\Lambda = 1-\Omega_{\rm m}$, and $\Omega_{\rm b}=0.0486$.

\begin{figure*}
\centering
\includegraphics[width=87mm]{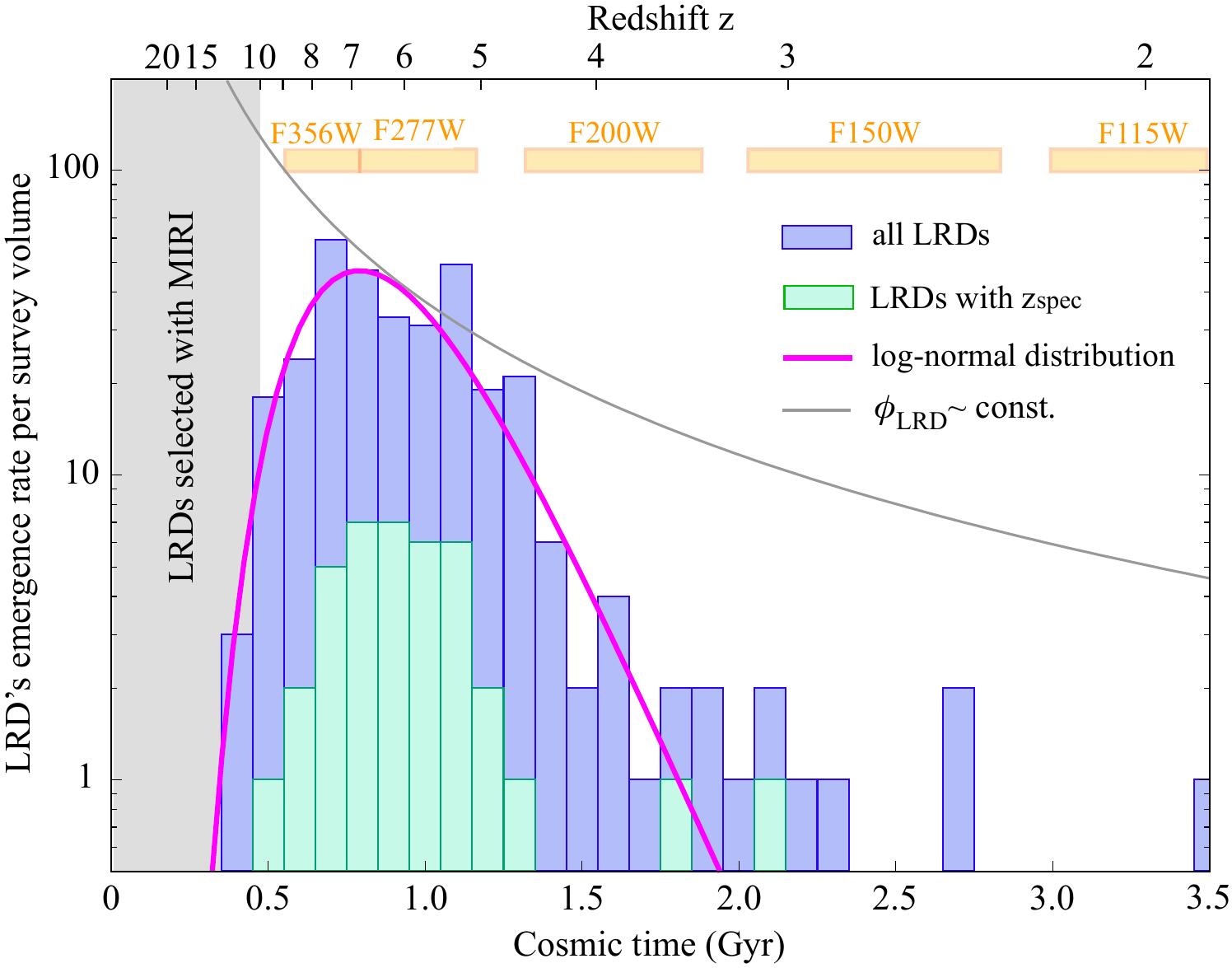}\hspace{5mm}
\includegraphics[width=87mm]{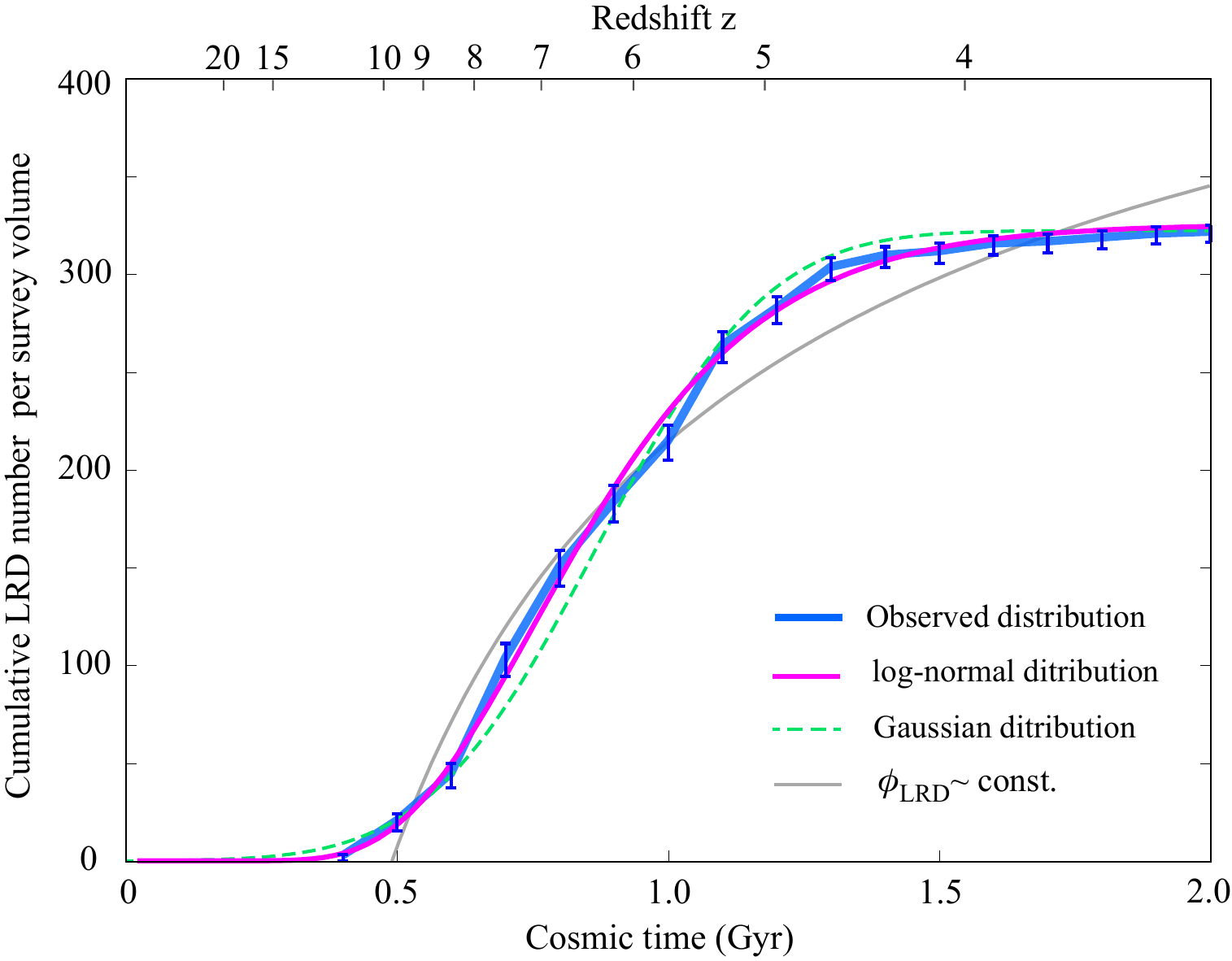}
\caption{{\it Left}: The occurrence rate of LRDs as a function of cosmic time (bottom $x$-axis) and redshift (top $x$-axis).
The histogram, based on the data from \cite{Kocevski_2025}, shows 341 photometrically-selected LRDs (blue), including 39 with
spectroscopically-confirmed redshifts (green).
The colored bars indicate the redshift range where the $4000~{\rm \AA}$ break falls within the bandpass of each wide-band filter.
The photometric selection of $z>9$ LRDs becomes incomplete (gray shaded region).
The magenta curve represents the best-fit function with a log-normal distribution, while the gray curve shows the case with constant abundance 
in unit comoving volume for reference.
{\it Right}: The cumulative number distribution of the observed LRD number (blue) with error bars computed via binomial statistics.
The best-fit result obtained using a log-normal distribution is shown ($p=0.873$), along with a Gaussian distribution yielding a worse fit ($p=1.17\times 10^{-4}$).
}
\label{fig:distribution}
\vspace{5mm}
\end{figure*}

\vspace{3mm}
\section{Cosmic Evolution of LRD Numbers}

To date, more than 300 LRDs have been identified through JWST survey programs, using photometric color selection techniques.
We adopt a sample of 341 LRDs spanning over $z\sim 2-11$, compiled from the CEERS, PRIMER, JADES, UNCOVER and NGDEEP surveys 
\citep{Kocevski_2025}.
These LRDs were selected using a spectral slope fitting technique, which employs shifting bandpasses to sample the same rest-frame 
emission both blueward and redward of the $4000~{\rm \AA}$ break, and enables a self-consistent search for AGN candidates with 
red optical and blue UV colors over a wide range of redshifts (see also \citealt{Hainline_2025}).

In the left panel of Figure~\ref{fig:distribution}, we present the occurrence rate of the LRD sample as a function of cosmic time (bottom $x$-axis) and redshift (top $x$-axis).
The histogram shows 341 photometrically-selected LRDs (blue), including 39 with spectroscopically-confirmed redshift.
The colored bars indicate the redshift range where the 4000~\AA\ break falls within the bandpass of each wide-band filter.
These LRDs with unique v-shape spectra and compact morphology emerge at $z\sim 8$, but the number begins to decline at $z\lesssim 6$, 
and experience a rapid drop at $z\sim 4$ \citep{Kocevski_2025}.
For $z>8$, the turnover wavelength enters the F356W band, while red optical continua are observed only in F444W (and F410M, if available).
As a result, the selection of $z>9$ LRDs becomes incomplete and thus the number might drop toward higher redshifts\footnote{
Recently, a $z\sim 9$ LRD candidate has been spectroscopically confirmed as a broad-line (H$\beta$) AGN at $z_{\rm spec}=9.288$ 
\citep[CAPERS-LRD-z9,][]{Taylor_2025b}. This object will be an anchor point to test the log-normal distribution hypothesis.}.
On the low-redshift side ($z < 4$), the number of LRD detections rapidly decreases, even though the available NIRCam bands are capable 
of capturing the $4000~{\rm \AA}$ break. 
It is worth noting that the histogram distributions of the total LRD sample and the subset with spectroscopic redshifts show similar shapes
without significant skewness.
This suggests that photometric redshift measurements provide a reliable estimate of their true values for these LRD samples.
For comparison, we consider a scenario where the comoving number density of LRDs remains constant over time, 
yielding a reference evolution for the occurrence rate: $\D N/\D t = \phi \cdot (\D V_{\rm c}/\D z)(\D z/\D t) \propto t^{-5/3}$, 
or equivalently $ \propto (1+z)^{5/2}$ (gray curve).
The observed decline in the LRD occurrence rate, however, deviates significantly from the reference evolution.

This implies that LRD activity is not a process that repeats continuously over cosmic time,
but instead occurs sporadically due to complex and random factors.
In many natural systems (e.g., earthquakes and astronomical phenomena such as gamma-ray bursts and fast radio bursts), the waiting time between successive events 
follows a log-normal distribution \citep[e.g.,][]{Li_Fenimore_1996,Ellsworth_1999,Kirsten_2024}, reflecting the stochastic nature of these events.
Inspired by these analogies, we employ the standard chi-square minimization technique to fit the observed data with errors estimated from Poisson statistics,
using a log-normal distribution defined as 
\begin{equation}
\frac{\D N}{\D t} =\frac{N_0}{\sqrt{2\pi} \sigma_0 t}\exp \left[ -\frac{\{\ln (t/t_0) \}^2}{2\sigma_0^2}\right].
\label{eq:logN}
\end{equation}
From the fitting procedure, we find  the best-fit parameters (reduced chi-square $\chi^2_\nu=2.02$): 
$t_0=865 \pm 22~{\rm Myr}$ and $\sigma_0=0.297\pm 0.0184$ with a time bin of $\Delta t=100~{\rm Myr}$ (magenta curve).
Importantly, the log-normal fit remains robust even when the histogram is restricted to LRDs with UV absolute magnitudes, 
$M_{\rm UV}<-18$ mag, where detection is not significantly affected by flux limits. 
This strengthens the validity of our argument, indicating that the observed trend is intrinsic to LRDs rather than a selection effect
(see more details in Appendix~\ref{sec:selection}).

The right panel of Figure~\ref{fig:distribution} shows the cumulative number distribution of LRDs.
For $t>1.3~{\rm Gyr}$, the distribution flattens due to the sharp decline in the number of sources.
For comparison, we overlay three curves representing different fitted distribution functions for the occurrence rate:
a log-normal distribution (magenta), a Gaussian distribution (green), and a power-law form with $\D N/\D t \propto t^{-5/3}$ (gray), which 
assumes a constant $\phi_{\rm LRD}$.
The log-normal distribution provides our best-fit model ($\chi^2_\nu = 0.647$ and $p=0.873$) with 
  $t_0= 837\pm 6~{\rm Myr}$ and $\sigma_0=0.327\pm 0.00803$.
These values are in good agreement with the previous fitting results for the occurrence rate\footnote{
Although the fitting result generally depends on the choice of time bin, the consistency of the fitted parameters based on 
the differential and cumulative distribution functions ensures the robustness of the fitting result.}.
The Gaussian model yields a slightly worse fit ($\chi^2_\nu = 2.65$ and $p=1.17\times 10^{-4}$), as the function form fails to capture the asymmetric feature of the distribution.
The low $p$-value ($\ll1$) indicates that the Gaussian model is statistically inconsistent with the observed data.
The power-law model does not explain the flattening of the cumulative distribution ($\chi^2_\nu = 13.6$). 
Even when the power-law index is treated as a free parameter, the best-fitting function ($\D N/\D t \propto t^{-2.85}$) still 
does not match the observed trend well.
Such a steep decline cannot be only attributed to the loss of LRD features, particularly their compact morphology, 
due to mergers with normal galaxies (see Appendix~\ref{sec:merger}), though merger may play a partial role in their disappearance
\citep{Khan_2025,Inayoshi_2025b}.

Assuming a log-normal form for the occurrence rate given in Equation~(\ref{eq:logN}), the cosmic evolution of LRDs is described by
the following expression
\begin{align}
\phi_{\rm LRD} &= \phi_0 f(z) \exp\left[  -\frac{\left\{\ln (1+z)-\mu_z \right\}^2}{2\sigma_z^2}\right],
\label{eq:fit1}
\end{align}
where $\log(\phi_0/{\rm Mpc}^{-3})=-5.2762\pm0.0015$,
$\mu_z=\ln(1+z_0)$ with a characteristic redshift of $z_0(=6.53^{+0.04}_{-0.03})$ corresponding to the cosmic age 
$t_0(= 837\pm 6~{\rm Myr})$, 
and $\sigma_z = 2\sigma_0/3(=0.218\pm 0.00535)$. Here, the cosmic age is converted to redshift using 
$t\simeq 2/(3H_0\sqrt{\Omega_{\rm m}})(1+z)^{-3/2}$, which holds for $z\gg (\Omega_{\Lambda}/\Omega_{\rm m})^{1/3}-1\simeq 0.3$.
The function $f(z)$ accounts for the redshift dependence of both the comoving volume element per redshift bin
${\rm d}V_{\rm c}/{\rm d}z$, and the cosmic time interval ${\rm d}t/{\rm d}z$, and is well approximated with the analytical form
\begin{equation}
f(z)=\frac{(1+z)^{3/2}}{[s(1+z)^{1/2}-1]^2},
\label{eq:fit2}
\end{equation}
and $s=0.903$ is fitted for a flat $\Lambda$CDM universe with $\Omega_{\rm m}=0.307$ at $z\geq 1.5$\footnote{
The functional form in Equation~(\ref{eq:fit2}) is motivated by the integral form of 
$\int^z_0 dz'/\sqrt{\Omega_{\rm m}(1+z')^3+\Omega_\Lambda}$ in the matter-dominant universe. 
The parameter $s$ is determined by the integral contribution from the low-redshift universe.
For $\Omega_{\rm m}=0.3$, a value commonly adopted in observational studies, we obtain $s=0.90066$.}.
These formulae in Equations~(\ref{eq:fit1}) and (\ref{eq:fit2}) with three parameters ($\phi_0$, $\mu_z$, and $\sigma_z$) 
provide an effective modeling for the abundance and redshift evolution of LRDs as new observational constraints 
become available, particularly at lower redshifts \citep[see][]{Ma_2025}.
This analytical form can also be applied to study the redshift evolution of LRD luminosity functions within individual magnitude bins.

\section{Discussion}

Based on our finding that the occurrence rate of LRDs follows a log-normal distribution, we propose a hypothesis for the origin of LRDs:
the first one or two AGN events associated with a newly-formed seed BH are observed as LRDs due to the unique characteristics of the surrounding environment.
After these initial accretion episodes, the LRD features fade and the objects transition into normal AGNs.

In this scenario, LRDs represent the earliest phase of AGN activity from newly-born BH seeds with $\sim 10^{4-5}~\msun$, which are likely embedded in
dense gas environments \citep[e.g.,][]{Bromm_Loeb_2003, Begelman_2006, Shang_2010} with abundances of $\sim 10^{-5}-10^{-4}~{\rm Mpc}^{-3}$,
corresponding to those of atomic-cooling halos in overdense regions where intense Lyman-Werner radiation and/or violent halo mergers 
enhance the possibility of BH seeding through suppression of H$_2$ formation and cooling \citep[e.g.,][]{Valiante_2016,Li_2021,Li_2023,Trinca_2022}.
This dense gas leads to strong absorption features on top of the broad Balmer emission lines, which is one of the key spectral signatures 
for LRDs \citep{Matthee_2024,Juodzbalis_2024,Ji_2025}.
Furthermore, if the gas surrounding the seed BH accretes at super-Eddington rates, this would naturally explain the observed 
X-ray weakness and weak (but observable) variability of LRDs \citep{Madau_Haardt_2024,Inayoshi_2025}. 
When the BH grows rapidly at a rate with an Eddington ratio of $\lambda_{\rm Edd} \sim O(10)$, the $e$-folding time of BH growth is 
$t_{\rm grow} \simeq 1.5~{\rm Myr}~(\lambda_{\rm Edd}/30)^{-1}$. 
This implies that the seed BH can reach a mass of $10^7~\msun$ within just one or two accretion episodes, each lasting several Myrs \citep[e.g.,][]{Inayoshi_2022a}. 
Once the BH grow in mass substantially, super-Eddington accretion becomes unsustainable due to a reduced mass supply from 
the galaxy environment \citep{Hu_2022b,Scoggins_2024,Hu_2025}. 
As a result, the BH transitions from the early super-Eddington accretion phase to a more typical AGN phase,
after making the BHs largely overmassive compared to the BH-to-galaxy mass ratio measured in the nearby universe
\citep[e.g.,][]{Harikane_2023_agn,Maiolino_2024_JADES,Chen_2024,Taylor_2025b,Kocevski_2025}.

\begin{figure*}
\centering
\includegraphics[width=87mm]{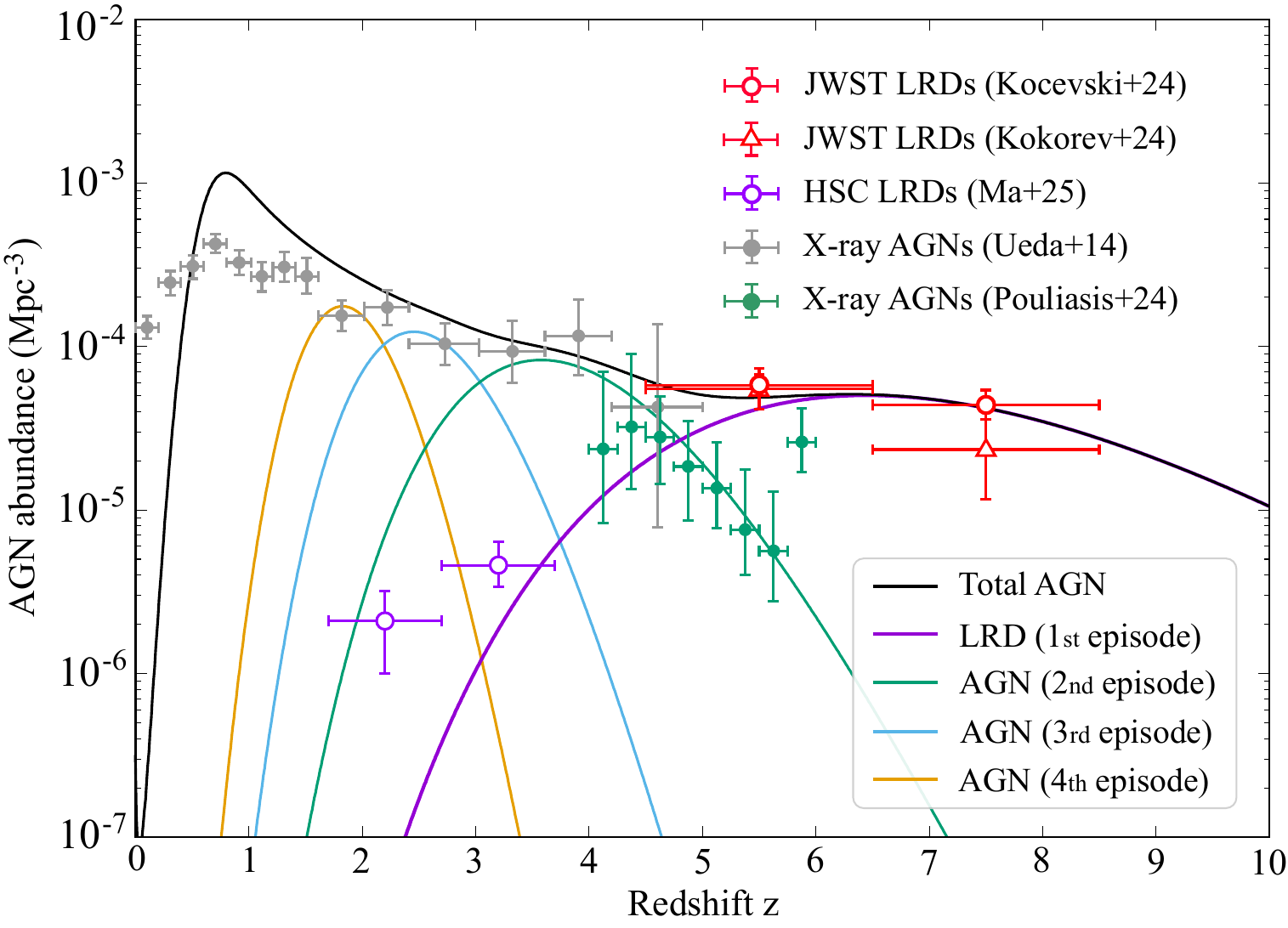}\hspace{3mm}
\includegraphics[width=87mm]{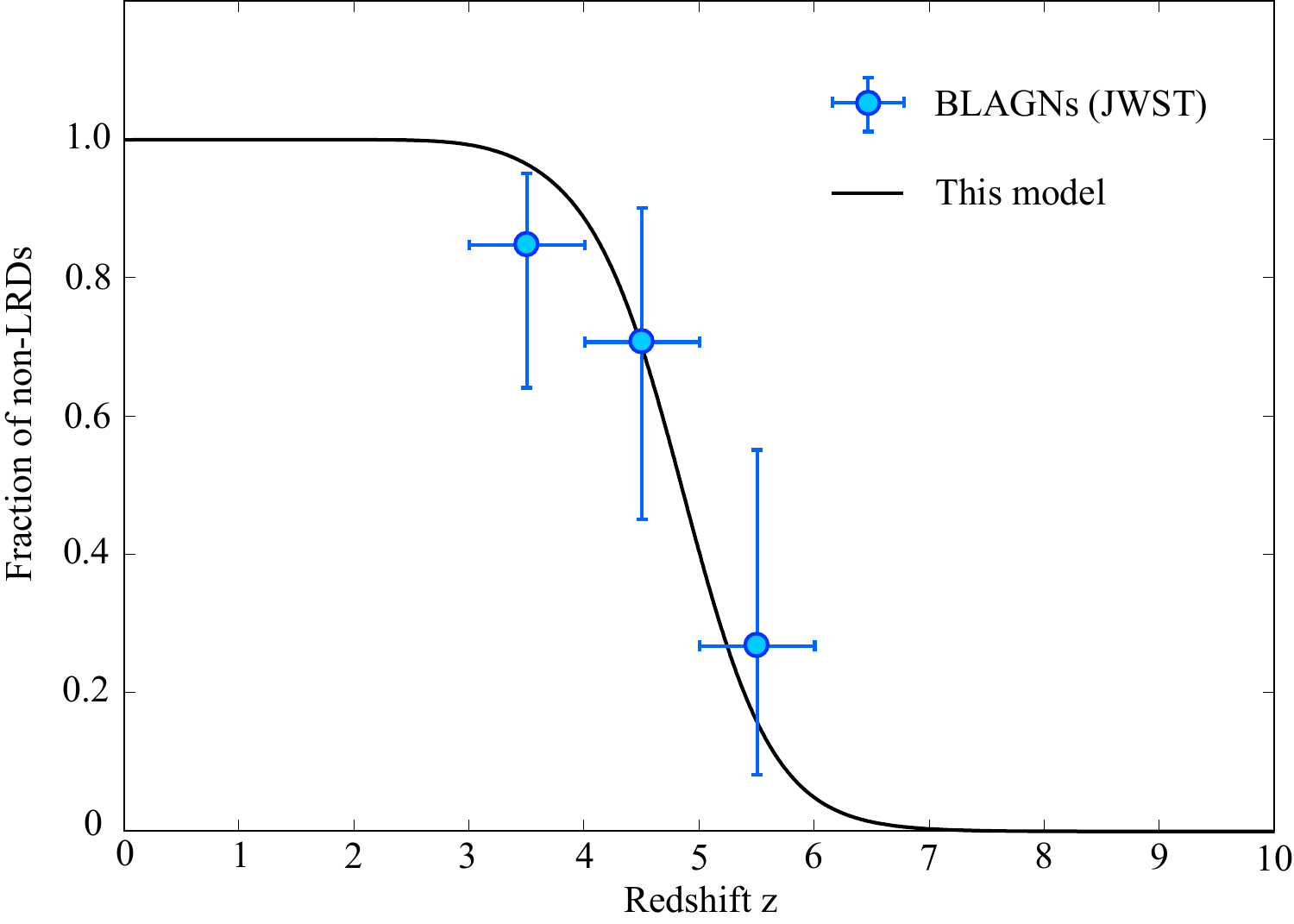}
\caption{{\it Left}: The redshift dependence of the AGN comoving number density.
Each modeled curve presents the AGN population that undergo $n$-th AGN active episodes, illustrating the distribution of the total elapsed time $T_n$ ($1\leq n\leq 8$): 
the first (magenta), second (green), third (cyan), fourth (orange) episodes, as well as the total population in this model (black). 
The normalization is set such that the modeled LRD abundance is consistent with the observed ones \citep[red,][]{Kokorev_2024a,Kocevski_2025}.
The predicted abundance in this model agrees with those of LRD candidates at $1.7<z<3.7$ photometrically selected 
by wide-area ground-based telescope surveys \citep[purple;][]{Ma_2025}, as well as with X-ray selected AGNs over $0<z<5$ 
\citep[gray;][]{Ueda_2014} and $4<z<6$ \citep[green;][]{Pouliasis_2024}.
{\it Right}: The fraction of non-LRDs (black curve; model) and observational constraints on the fraction (95\% confidence level) derived from a sample of 
62 broad-line AGNs \citep[blue symbol;][]{Taylor_2025a}.
}
\label{fig:phi}
\vspace{5mm}
\end{figure*}

Assuming the physical processes driving both LRDs and normal AGNs are essentially the same, and the time interval between 
each individual event follows a similar statistical distribution, the cosmic evolution of AGN abundances should evolve in a similar fashion for LRDs.
However, observations reveal that while LRDs show a rapid decline in occurrence rate, the occurrence rate of normal AGNs remains nearly constant 
down to low redshifts \citep{Ueda_2014,Delvecchio_2014,Pouliasis_2024}.
This discrepancy suggests that factors beyond the timing of single events (for instance, the recurrence of AGN activity)
play a role in shaping the cosmic evolution of the bulk AGN population.

To investigate this, we model the occurrence times of AGN events as independently and identically distributed, making two key assumptions:
(1) the occurrence time follows a single distribution in each episode, and
(2) the occurrence time for each individual AGN is independent of others, with no direct influence on the timing of subsequent events.
Under these assumptions, we generate a set of independent random variables $t_i$, drawn from a log-normal distribution to represent individual AGN occurrence times.
We then compute the probability distribution of the total elapsed time across multiple episodes, defined as the sum of 
$n$ independent log-normal variables,  $T_n=\sum_{i=1}^n t_i$.
To obtain a statistically robust distribution, we perform Monte Carlo sampling for each $T_n$ and numerically construct its probability distribution.
Note that as $n\rightarrow \infty$, the distribution of $T_n$ approaches a Gaussian distribution according to the central limit theorem,
allowing for an analytical treatment for this calculation (see Appendix~\ref{sec:analytical}).

In the left panel of Figure~\ref{fig:phi}, each curve presents the AGN population categorized by the number of active episodes that the nuclear BH undergo,
showing the distribution of $T_n$: the first ($n=1$; magenta), second ($n=2$; green), third ($n=3$; cyan), fourth ($n=4$; orange) episodes, as well as the total population,
which includes up to eight episodes ($n=8$; black). 
The normalization of the modeled abundance is set such that the number of objects in their first active episodes matches the observed abundance of LRDs
at $z\sim 4-7$, which is $\phi_{\rm LRD}\simeq (2-5)\times 10^{-5}~{\rm Mpc}^{-3}$ integrated over $-20\leq M_{\rm UV}/{\rm mag}\leq -18$ \citep{Kokorev_2024a,Kocevski_2025}.
Since the total number of objects is conserved across successive episodes, the modeled AGN abundance at $z<5$ is not explicitly fitted to the observational data.
Nevertheless, the predicted abundance from this log-normal model agrees well with observations of LRD candidates at lower redshifts, photometrically selected 
by wide-area ground-based telescope surveys \citep{Ma_2025}, as well as with X-ray selected AGNs over $0<z<5$ \citep{Ueda_2014} and $4<z<6$ \citep{Pouliasis_2024}.
As the LRD number decreases at $z < 4$, the total AGN abundance curve increases toward lower redshifts,
reflecting the shift from LRDs to more typical AGN populations.
This trend is consistent with our finding of $\phi \propto (1+z)^{-5/2}$,
which is derived from the nearly constant occurrence rate of AGN activity (see Equation~\ref{eq:conver}). 
Moreover, the rapid emergence of X-ray selected AGNs observed at $4<z<6$ aligns well with the prediction of this hypothesis, as shown by the green curve 
indicating AGNs that undergo their second episodes.
This agreement further supports the idea that LRDs represent the earliest phase of AGN activity, after which the objects evolve into normal AGNs.

In the right panel of Figure~\ref{fig:phi}, we show the fraction of AGNs that have experienced their second or later episodes relative to the total
AGN population, representing the fraction of non-LRDs under this hypothesis.
Since a significant fraction of BHs undergo a second episode at $z<5$, this fraction increases rapidly and reaches unity by $z\sim 3$.
For comparison, we overlay observational constraints on the fraction of unobscured (non-LRD) AGNs (95\% confidence level), 
derived from a sample of 62 spectroscopically confirmed broad-line AGNs in the redshift range of $3.5<z<6$ \citep{Taylor_2025a}.
At $z\leq 6$, the sample size in each redshift bin is sufficiently large to reveal a trend that closely follows the model prediction (black curve)\footnote{
Since the sample size at $6<z<7$ is too small for robust statistical analysis (one LRD and three unobscured AGNs),
we do not include the data in the right panel of Figure~\ref{fig:phi}. Statistical uncertainties are evaluated using the Clopper‐Pearson method.}.
The redshift range where LRDs begin to diminish coincides with the emergence of typical unobscured AGNs.

The emergence conditions of LRDs are likely tied to the underlying BH seeding mechanisms and surrounding environment.
One proposed model is the heavy seed scenario, in which seed BHs form through the direct collapse of massive atomic gas clouds 
(e.g., \citealt{Loeb_Rasio_1994,Bromm_Loeb_2003,Begelman_2006,Shang_2010}, see also \citealt{Inayoshi_ARAA_2020}).
Alternatively, the lack of an obvious host galaxy component in LRDs \citep{Chen_2024} implies that processes beyond baryonic 
physics might be involved.
For instance, the collapse of a halo core via energy dissipation in a self-interacting dark matter environment has been proposed
\citep[e.g.,][]{Feng_2021,GrantRoberts_2025,Jiang_2025}. 
In both scenarios, a key prediction is that BH formation precedes active star formation
and yields a BH-to-stellar mass ratio substantially higher than the local empirical value \citep[e.g.,][]{Hu_2025,Jiang_2025}.
Notably, a recently reported LRD with its broad H$\beta$ emission at $z_{\rm spec} = 9.288$, CAPERS-LRD-z9, shows an apparent overmassiveness in the ratio \citep{Taylor_2025b}, offering opportunity for testing the BH seeding channel along with
the transitional trend of LRD abundance relative to the entire BH population (see the right panel of Figure~\ref{fig:phi}).
Moreover, the initial accretion episodes onto newly-born seed BHs are likely led by intense cold flows along cosmic filaments
in overdense regions of the universe \citep{DiMatteo_2012} and may occur through gas with low angular momentum \citep{Eisenstein_Loeb_1995}. 
The timing of these accretion processes, potentially influenced by the spin distribution of the halo, may be crucial for the emergence of LRDs
(see also \citealt{Kroupa_2020}).

The physical origin of the log-normal occurrence distribution for LRDs (and more broadly, AGN activity) remains an open question.
However, the observed statistical properties suggest that the occurrence time, $t_{\rm AGN}$, is determined by the product of multiple 
independent physical quantities $x_i$ ($i=1,2,\cdots n$) with some inherent randomness, such that
\begin{equation}
t_{\rm AGN} = x_1 x_2\cdots x_n. 
\end{equation}
If the logarithms of these variables, $\log x_i$, follow distributions that satisfy weak conditions, the central limit theorem implies that the distribution of
$\log t_{\rm AGN}$ tends toward a normal distribution as $n$ increases.
In astrophysical contexts, while $n \rightarrow \infty$ is not strictly applicable, even a modest number ($n\sim 3$) is sufficient to produce a log-normal distribution 
\citep[e.g.,][]{Ioka_Nakamura_2002}.
This statistical (i.e., {\it macroscopic}) property can be described without specifying the detailed {\it microscopic} physical processes that govern AGN triggering.
Modeling the specific mechanisms that set the mean and variance of the distribution will be left for future studies 
\citep[e.g., BH coalescences via gravitational-wave emission,][]{Inayoshi_2025b}.

\newpage

\acknowledgments
I greatly thank Changhao Chen, Kejian Chen, Zolt\'an Haiman, Luis C.~Ho, Kenta Hotokezaka, Kohei Ichikawa, Masafusa Onoue,
and Jinyi Shangguan for constructive discussions, and Anthony J.~Taylor, Dale D.~Kocevski, and Ektoras Pouliasis for kindly providing data and granting 
permission for its use in this study.
I acknowledges support from the National Natural Science Foundation of China (12233001), 
the National Key R\&D Program of China (2022YFF0503401), and the China Manned Space Program (CMS-CSST-2025-A09).
This work is based on observations made with the NASA/ESA/CSA James Webb Space Telescope. 
The data were obtained from the Mikulski Archive for Space Telescopes at the Space Telescope Science Institute, which is operated by the Association of Universities for Research in Astronomy, Inc., under NASA contract NAS 5-03127 for JWST.

\appendix

\section{Selection}\label{sec:selection}

\begin{figure*}
\centering
\includegraphics[width=83mm]{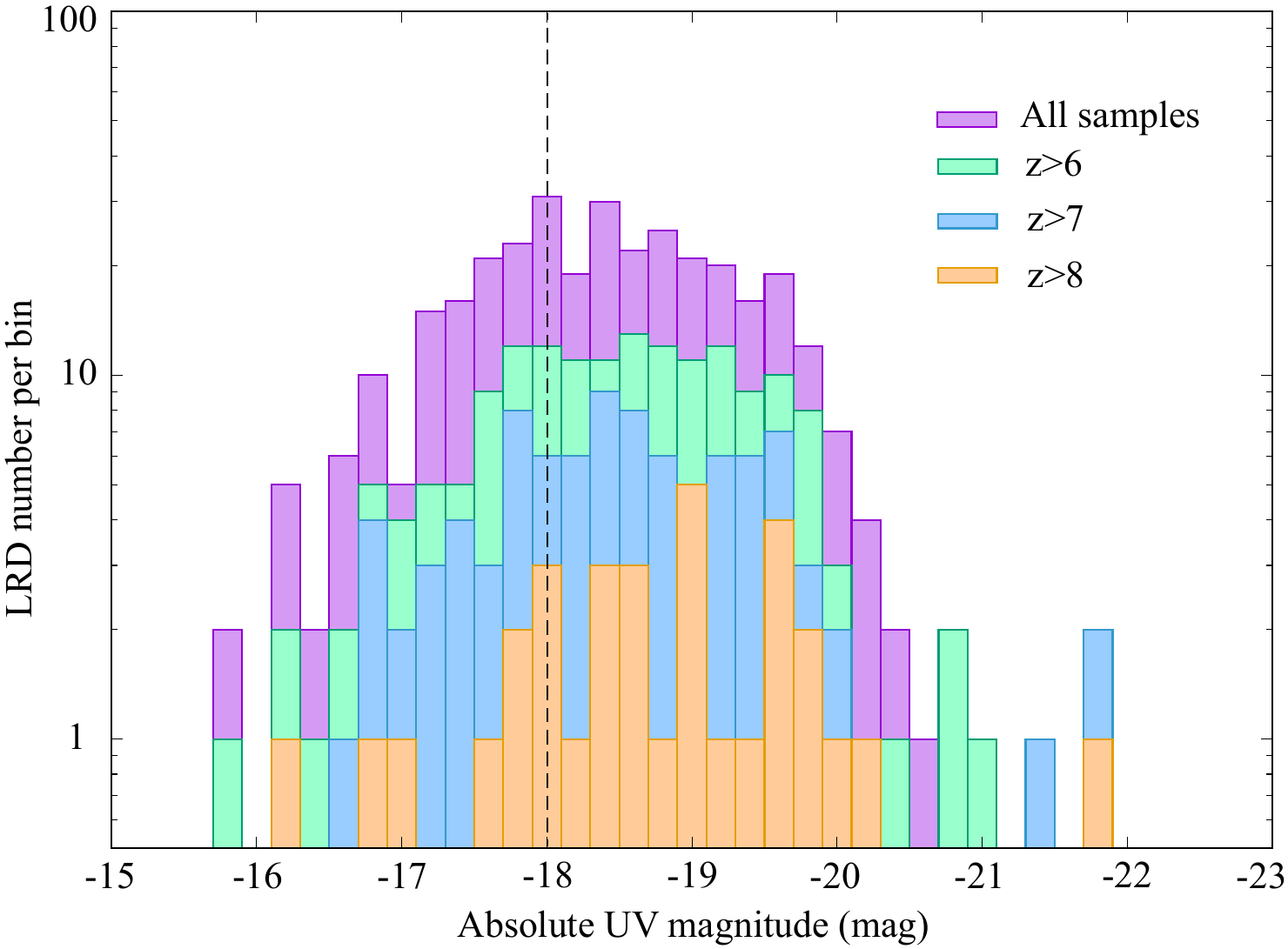}\hspace{5mm}
\includegraphics[width=83mm]{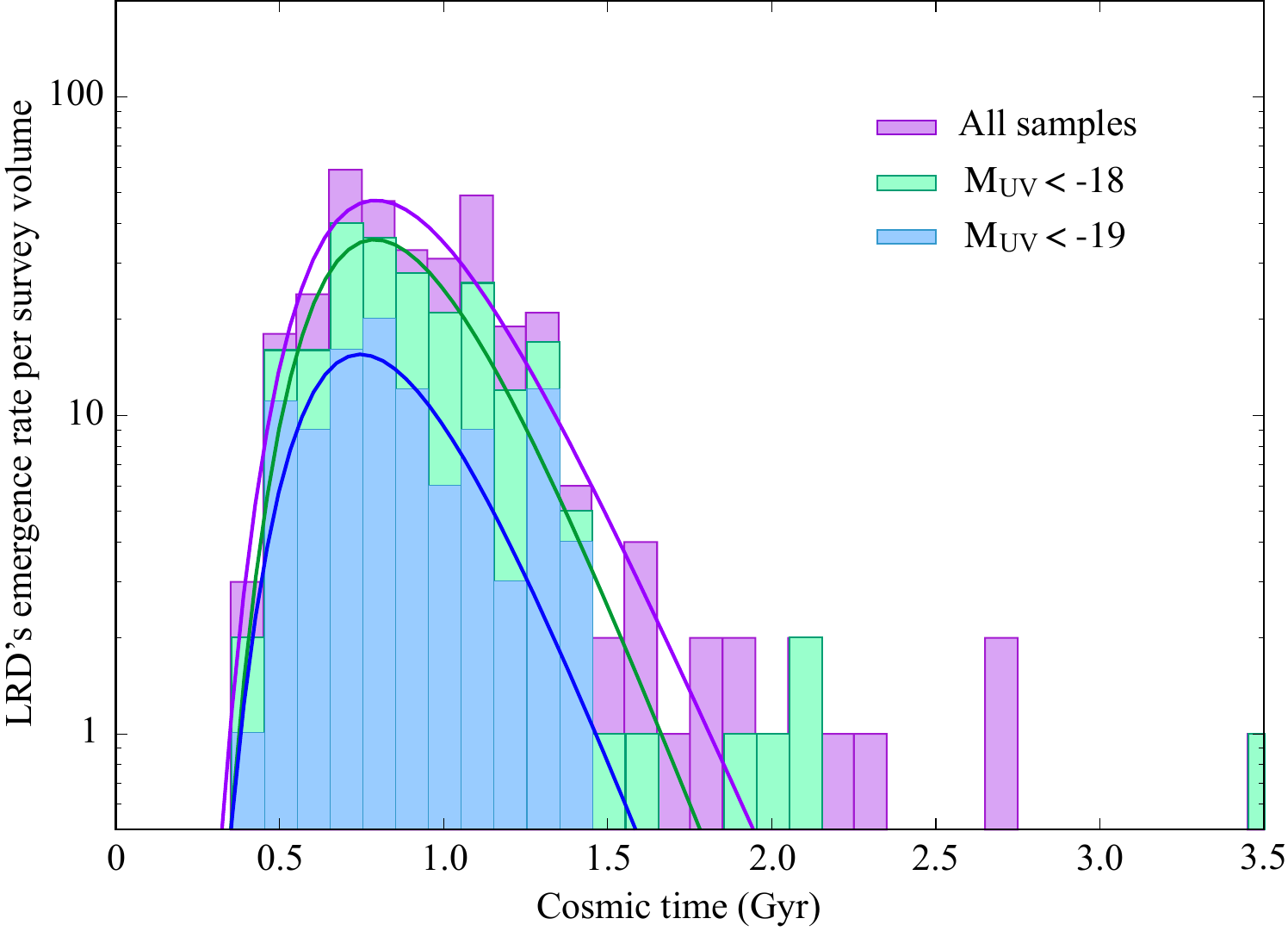}
\caption{{\it Left}: The distribution of UV absolute magnitudes for LRDs with the total sample (purple) and subsets at $z\geq 6$ (green), 
$z\geq 7$ (blue), and $z\geq 8$ (orange). The vertical line indicates a reference threshold of $M_{\rm UV}=-18$ mag,
fainter than which the observed decline in number is likely due to flux limits.
{\it Right}: The LRD occurrence rate for the total sample (purple) and subsets at $M_{\rm UV}<-18$ (green) and $M_{\rm UV}<-19$ (blue)
overlaid with best-fit log-normal distributions (solid curves).
}
\label{fig:Muv}
\vspace{8mm}
\end{figure*}

The left panel of Figure~\ref{fig:Muv} shows the distribution of LRD detections as a function of absolute UV magnitude
with the total sample (purple) and subsets at $z\geq 6$ (green), $z\geq 7$ (blue), and $z\geq 8$ (orange).
The histogram peaks at $M_{\rm UV}\sim -18$ mag with a decline on both the brighter and fainted sides.
The rarity of LRDs on the brighter end suggests an intrinsically lower abundance, while the decline on the fainter end is likely due to flux limits, 
as also seen in the UV luminosity functions at $M_{\rm UV}>-18$ mag \citep{Kocevski_2025}.
The characteristic bending magnitude remains consistent across all redshift bins.

The right panel of Figure~\ref{fig:Muv} presents the LRD occurrence rate for each UV-magnitude bin, along with the best-fit log-normal distribution (solid curves).
The fitted parameters, summarized in Table~\ref{tab:fit}, fall within the $1\sigma$ uncertainty range.
Even when considering only bright LRDs with $M_{\rm UV}<-18$ mag, the emergence trend of LRDs at $t\gtrsim 0.5$ Gyr remains robust.
Thus, we conclude that this trend reflects an intrinsic phenomenon rather than a selection effect.

\vspace{2mm}
\section{Diminishing LRD features due to mergers with other galaxies}\label{sec:merger}
\begin{figure}
\centering
\includegraphics[width=83mm]{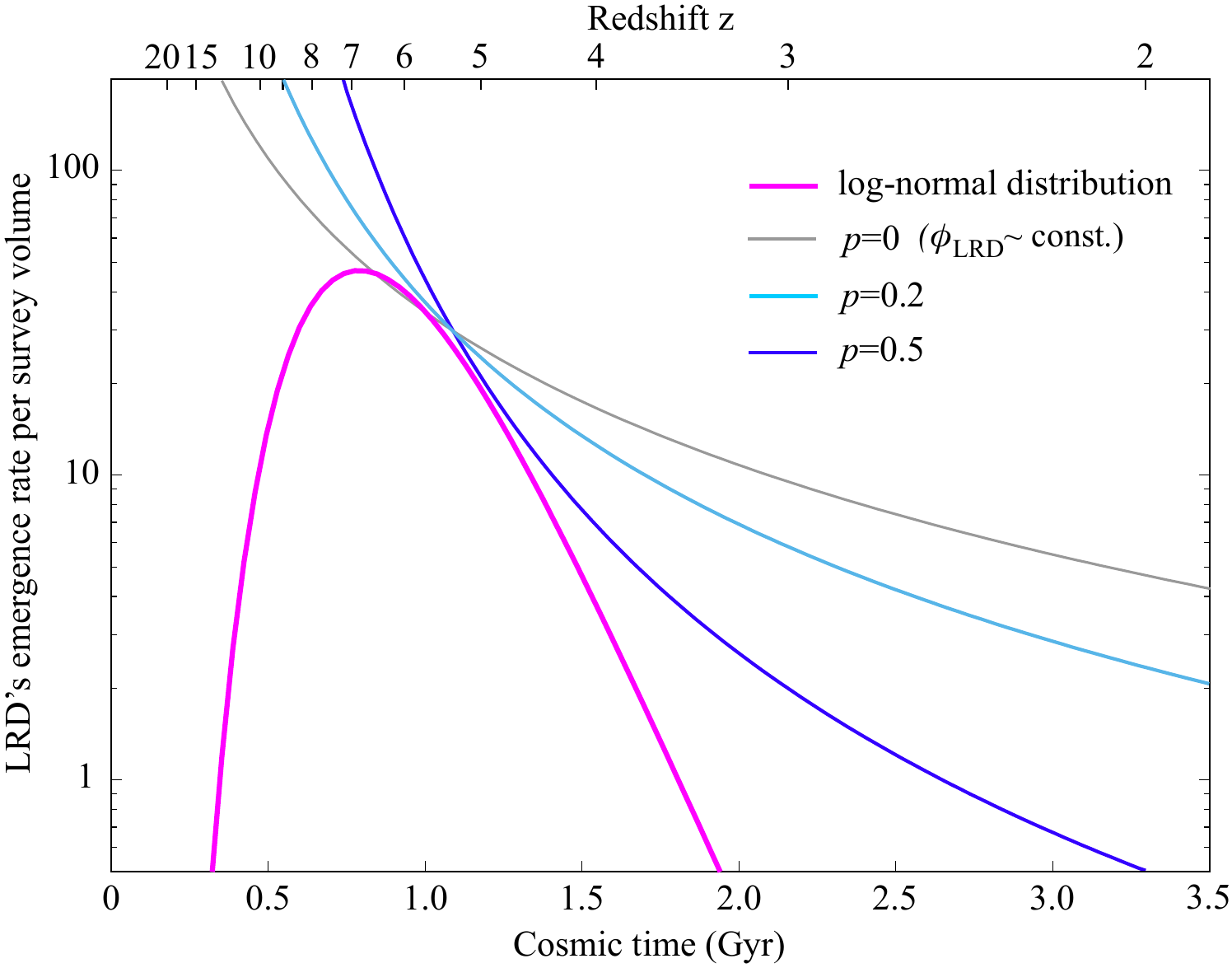}
\caption{The LRD occurrence-rate models: the log-normal case (magenta) and the decay cases driven by mergers: $p=0$ (gray), $p=0.2$ (cyan), and $p=0.5$ (blue).
}
\label{fig:mrg}
\vspace{2mm}
\end{figure}

One possible explanation for the decline in the LRD occurrence rate at $z<4-5$ is that their unique features, especially their compact morphology \citep[e.g.,][]{Labbe_2025,Kocevski_2025}, 
are lost due to mergers with normal galaxies. 
Major mergers could cause LRDs to appear as extended sources, blending stellar components with the underlying AGN and diminishing their LRD-like characteristics.
Indeed, dual LRD candidates with $\sim 1-2$ kpc projected separations have been reported \citep{Tanaka_2024b}.
Cosmological $N$-body simulations estimate that major mergers (with a mass ratio of $\xi \gtrsim 0.3$) occur with a frequency of $p \sim 0.2$ per halo per redshift interval \citep{Fakhouri_2010}.
This suggests that the number of LRDs that avoid such mergers and retain their unique features decreases toward lower redshifts as
\begin{equation}
\frac{\D N_{\rm LRD}}{\D z} \simeq N_0 e^{-p(z_0-z)}.
\end{equation}
Figure~\ref{fig:mrg} presents two curves representing $\D N_{\rm LRD}/\D t$ for $p = 0.2$ (cyan) and $p=0.5$ (blue), 
where $N_0$ is adjusted to match the histogram at a cosmic age of $\sim 1$ Gyr. 
The case with $p=0.2$ approximately corresponds to a scenario where the comoving number density of LRDs follows a power-law evolution, $\phi \propto t^{-1}$,
consistent with the overall cosmic evolution of the number density of galaxies over $0<z<8$ \citep{Conselice_2016}.
Even in an extreme case where $p=0.5$ (implying that 50\% of LRDs undergo merger in each redshift interval, a rate observed in simulations for minor mergers),
the resulting decline trend is still significantly shallower than the log-normal decay (magenta curve).
This suggests that mergers alone cannot fully account for the rapid disappearance of LRDs at lower redshifts,
unless minor mergers with $\xi\gtrsim 0.03$ would diminish LRD characteristics.

In addition, it is worth noting that the mass growth of LRD host galaxies through subsequent halo and galaxy mergers is 
likely a key driver of their evolution toward other galaxy populations.
In particular, gas-rich mergers can trigger intense starbursts while maintaining compact morphologies \citep{Oser_2010}, 
potentially transforming LRDs into ``red nuggets" by intermediate redshifts---systems considered to be the progenitors of present-day 
giant elliptical galaxies \citep{Thomas_2005,Trujillo_2009}.
An intriguing property of an LRD has been reported by \citet{Kokorev_2024b}, which exhibits a spectrum dominated by evolved stellar populations
with a mass of $M_\star\sim 10^{10.6}~\msun$ and no signs of ongoing star formation.
Other studies have similarly identified massive quiescent galaxies at high redshifts \citep[e.g.,][]{Carnall_2023,Wang_2024b,Onoue_2024,deGraaff_2025}, 
whose cosmic abundance approaches the upper limit predicted by the flat $\Lambda$CDM cosmology 
\citep[e.g.,][]{Carnall_2024}, as observed for LRD populations under an assumption that all observed light originates 
from stellar populations \citep{Wang_2024b,Inayoshi_Ichikawa_2024,Akins_2024}.
Importantly, the detection of broad H$\alpha$ emission lines in some of these quiescent galaxies
suggests that their star formation was quenched by AGN activity a few hundred million years prior to the observed epoch.
Future work will clarify the evolutionary connection between LRDs and their descendants at lower redshifts, 
as well as the transition from LRDs to more typical AGN populations.

\begin{table}[t]
\renewcommand\thetable{2} 
\caption{Fitting results with a log-normal distribution.}\vspace{-0mm}
\begin{center}
\begin{tabular}{ccc}
\hline
\hline
(1) & (2) & (3)\\
 & $t_0$ (Myr) & $\sigma_0$ \\
\hline 
All sample               & $865\pm20$  & $0.297\pm0.0184$ \\
$M_{\rm UV}<-18$  & $850\pm19$  & $0.278\pm0.0167$\\
$M_{\rm UV}<-19$  & $808\pm27$ &  $0.288\pm0.0258$\\
\hline 
\end{tabular}
\label{tab:fit}
\end{center}
\end{table}

\vspace{5mm}
\section{Analytic formula}\label{sec:analytical}

When the occurrence times of AGN events are independently and identically distributed, 
the central limit theorem states that the cumulative occurrence time until the $n$-th AGN event will 
approximately follow a Gaussian distribution with mean $n \mu$ and variance $n \sigma^2$, 
where $\mu$ and $\sigma^2$ are the mean and variance of the occurrence time distribution for each single event.
Although the occurrence timing follows a log-normal distribution, for simplicity, we approximate it as a Gaussian distribution with mean $\mu$ and variance $\sigma^2$. 
This assumption allows for an analytical treatment for the calculation of the cosmic AGN occurrence rate.
The total occurrence rate at a given cosmic time $t$ is expressed as
\begin{equation}
\frac{\D N_{\rm AGN} }{\D t}= \sum_{k=1}^n \frac{N_0}{\sqrt{2 \pi} \sigma_i}e^{-(t-\mu_k)^2/2\sigma_k^2},
\label{eq:sum}
\end{equation}
where $\mu_k= k \mu$ and $\sigma_k = \sqrt{k}\sigma$ for $1\leq k\leq n$.
For sufficiently large $k\gtrsim k_0\equiv(\mu/\sigma)^2$, the dispersion $\sigma_k$ becomes large enough that 
the separation between each Gaussian peak becomes smaller than the dispersion, and thus each Gaussian distribution largely overlaps.
Therefore, at later times of $t>k_0\mu =\mu^3/\sigma^2$, the sum can be approximated by an integral form and evaluated
using the saddle-point method around $x=t/\mu$,
\begin{align}
\frac{\D N_{\rm AGN}}{\D t} = \frac{N_0}{\sqrt{2 \pi} \sigma}\int_0^\infty \exp \left[\frac{-(t-\mu x)^2}{2\sigma^2x}\right] \frac{dx}{x}
\simeq \frac{N_0}{\mu}.
\label{eq:conver}
\end{align}
The result turns out {\it time-independent}.
This trend is consistent with the numerical result of $\phi\propto (1+z)^{-5/2}$ shown in Figure~\ref{fig:phi}, and aligns with measurements of 
AGN abundances in the range $0 < z < 5$ \citep{Ueda_2014} as well.

\bibliographystyle{aasjournal}
\bibliography{ref.bib}

\begin{thebibliography}{}
\expandafter\ifx\csname natexlab\endcsname\relax\def\natexlab#1{#1}\fi
\providecommand{\url}[1]{\href{#1}{#1}}
\providecommand{\dodoi}[1]{doi:~\href{http://doi.org/#1}{\nolinkurl{#1}}}
\providecommand{\doeprint}[1]{\href{http://ascl.net/#1}{\nolinkurl{http://ascl.net/#1}}}
\providecommand{\doarXiv}[1]{\href{https://arxiv.org/abs/#1}{\nolinkurl{https://arxiv.org/abs/#1}}}

\bibitem[{{Akins} {et~al.}(2024){Akins}, {Casey}, {Lambrides}, {Allen},
  {Andika}, {Brinch}, {Champagne}, {Cooper}, {Ding}, {Drakos}, {Faisst},
  {Finkelstein}, {Franco}, {Fujimoto}, {Gentile}, {Gillman}, {Gozaliasl},
  {Harish}, {Hayward}, {Hirschmann}, {Ilbert}, {Kartaltepe}, {Kocevski},
  {Koekemoer}, {Kokorev}, {Liu}, {Long}, {McCracken}, {McKinney}, {Onoue},
  {Paquereau}, {Renzini}, {Rhodes}, {Robertson}, {Shuntov}, {Silverman},
  {Tanaka}, {Toft}, {Trakhtenbrot}, {Valentino}, \& {Zavala}}]{Akins_2024}
{Akins}, H.~B., {Casey}, C.~M., {Lambrides}, E., {et~al.} 2024, arXiv e-prints,
  arXiv:2406.10341, \dodoi{10.48550/arXiv.2406.10341}

\bibitem[{{Begelman} {et~al.}(2006){Begelman}, {Volonteri}, \&
  {Rees}}]{Begelman_2006}
{Begelman}, M.~C., {Volonteri}, M., \& {Rees}, M.~J. 2006, \mnras, 370, 289,
  \dodoi{10.1111/j.1365-2966.2006.10467.x}

\bibitem[{{Bromm} \& {Loeb}(2003)}]{Bromm_Loeb_2003}
{Bromm}, V., \& {Loeb}, A. 2003, \apj, 596, 34, \dodoi{10.1086/377529}

\bibitem[{{Carnall} {et~al.}(2023){Carnall}, {McLure}, {Dunlop}, {McLeod},
  {Wild}, {Cullen}, {Magee}, {Begley}, {Cimatti}, {Donnan}, {Hamadouche},
  {Jewell}, \& {Walker}}]{Carnall_2023}
{Carnall}, A.~C., {McLure}, R.~J., {Dunlop}, J.~S., {et~al.} 2023, \nat, 619,
  716, \dodoi{10.1038/s41586-023-06158-6}

\bibitem[{{Carnall} {et~al.}(2024){Carnall}, {Cullen}, {McLure}, {McLeod},
  {Begley}, {Donnan}, {Dunlop}, {Shapley}, {Rowlands}, {Almaini},
  {Arellano-C{\'o}rdova}, {Barrufet}, {Cimatti}, {Ellis}, {Grogin},
  {Hamadouche}, {Illingworth}, {Koekemoer}, {Leung}, {Lovell},
  {P{\'e}rez-Gonz{\'a}lez}, {Santini}, {Stanton}, \& {Wild}}]{Carnall_2024}
{Carnall}, A.~C., {Cullen}, F., {McLure}, R.~J., {et~al.} 2024, arXiv e-prints,
  arXiv:2405.02242, \dodoi{10.48550/arXiv.2405.02242}

\bibitem[{{Chen} {et~al.}(2024){Chen}, {Ho}, {Li}, \& {Zhuang}}]{Chen_2024}
{Chen}, C.-H., {Ho}, L.~C., {Li}, R., \& {Zhuang}, M.-Y. 2024, arXiv e-prints,
  arXiv:2411.04446, \dodoi{10.48550/arXiv.2411.04446}

\bibitem[{{Conselice} {et~al.}(2016){Conselice}, {Wilkinson}, {Duncan}, \&
  {Mortlock}}]{Conselice_2016}
{Conselice}, C.~J., {Wilkinson}, A., {Duncan}, K., \& {Mortlock}, A. 2016,
  \apj, 830, 83, \dodoi{10.3847/0004-637X/830/2/83}

\bibitem[{{de Graaff} {et~al.}(2025){de Graaff}, {Setton}, {Brammer}, {Cutler},
  {Suess}, {Labb{\'e}}, {Leja}, {Weibel}, {Maseda}, {Whitaker}, {Bezanson},
  {Boogaard}, {Cleri}, {De Lucia}, {Franx}, {Greene}, {Hirschmann}, {Matthee},
  {McConachie}, {Naidu}, {Oesch}, {Price}, {Rix}, {Valentino}, {Wang}, \&
  {Williams}}]{deGraaff_2025}
{de Graaff}, A., {Setton}, D.~J., {Brammer}, G., {et~al.} 2025, Nature
  Astronomy, 9, 280, \dodoi{10.1038/s41550-024-02424-3}

\bibitem[{{Delvecchio} {et~al.}(2014){Delvecchio}, {Gruppioni}, {Pozzi},
  {Berta}, {Zamorani}, {Cimatti}, {Lutz}, {Scott}, {Vignali}, {Cresci},
  {Feltre}, {Cooray}, {Vaccari}, {Fritz}, {Le Floc'h}, {Magnelli}, {Popesso},
  {Oliver}, {Bock}, {Carollo}, {Contini}, {Le F{\'e}vre}, {Lilly}, {Mainieri},
  {Renzini}, \& {Scodeggio}}]{Delvecchio_2014}
{Delvecchio}, I., {Gruppioni}, C., {Pozzi}, F., {et~al.} 2014, \mnras, 439,
  2736, \dodoi{10.1093/mnras/stu130}

\bibitem[{{Di Matteo} {et~al.}(2012){Di Matteo}, {Khandai}, {DeGraf}, {Feng},
  {Croft}, {Lopez}, \& {Springel}}]{DiMatteo_2012}
{Di Matteo}, T., {Khandai}, N., {DeGraf}, C., {et~al.} 2012, \apjl, 745, L29,
  \dodoi{10.1088/2041-8205/745/2/L29}

\bibitem[{{Eisenstein} \& {Loeb}(1995)}]{Eisenstein_Loeb_1995}
{Eisenstein}, D.~J., \& {Loeb}, A. 1995, \apj, 443, 11, \dodoi{10.1086/175498}

\bibitem[{Ellsworth {et~al.}(1999)Ellsworth, Matthews, Nadeau, Nishenko,
  Reasenberg, \& Simpson}]{Ellsworth_1999}
Ellsworth, W.~L., Matthews, M.~V., Nadeau, R.~M., {et~al.} 1999, A
  physically-based earthquake recurrence model for estimation of long-term
  earthquake probabilities, Tech. rep., US Geological Survey

\bibitem[{{Fakhouri} {et~al.}(2010){Fakhouri}, {Ma}, \&
  {Boylan-Kolchin}}]{Fakhouri_2010}
{Fakhouri}, O., {Ma}, C.-P., \& {Boylan-Kolchin}, M. 2010, \mnras, 406, 2267,
  \dodoi{10.1111/j.1365-2966.2010.16859.x}

\bibitem[{{Feng} {et~al.}(2021){Feng}, {Yu}, \& {Zhong}}]{Feng_2021}
{Feng}, W.-X., {Yu}, H.-B., \& {Zhong}, Y.-M. 2021, \apjl, 914, L26,
  \dodoi{10.3847/2041-8213/ac04b0}

\bibitem[{{Grant Roberts} {et~al.}(2025){Grant Roberts}, {Braff}, {Garg},
  {Profumo}, {Jeltema}, \& {O'Donnell}}]{GrantRoberts_2025}
{Grant Roberts}, M., {Braff}, L., {Garg}, A., {et~al.} 2025, \jcap, 2025, 060,
  \dodoi{10.1088/1475-7516/2025/01/060}

\bibitem[{{Greene} {et~al.}(2024){Greene}, {Labbe}, {Goulding}, {Furtak},
  {Chemerynska}, {Kokorev}, {Dayal}, {Volonteri}, {Williams}, {Wang}, {Setton},
  {Burgasser}, {Bezanson}, {Atek}, {Brammer}, {Cutler}, {Feldmann}, {Fujimoto},
  {Glazebrook}, {de Graaff}, {Khullar}, {Leja}, {Marchesini}, {Maseda},
  {Matthee}, {Miller}, {Naidu}, {Nanayakkara}, {Oesch}, {Pan}, {Papovich},
  {Price}, {van Dokkum}, {Weaver}, {Whitaker}, \& {Zitrin}}]{Greene_2024}
{Greene}, J.~E., {Labbe}, I., {Goulding}, A.~D., {et~al.} 2024, \apj, 964, 39,
  \dodoi{10.3847/1538-4357/ad1e5f}

\bibitem[{{Hainline} {et~al.}(2025){Hainline}, {Maiolino}, {Juod{\v{z}}balis},
  {Scholtz}, {{\"U}bler}, {D'Eugenio}, {Helton}, {Sun}, {Sun}, {Robertson},
  {Tacchella}, {Bunker}, {Carniani}, {Charlot}, {Curtis-Lake}, {Egami},
  {Johnson}, {Lin}, {Lyu}, {P{\'e}rez-Gonz{\'a}lez}, {Rinaldi}, {Silcock},
  {Venturi}, {Williams}, {Willmer}, {Willott}, {Zhang}, \&
  {Zhu}}]{Hainline_2025}
{Hainline}, K.~N., {Maiolino}, R., {Juod{\v{z}}balis}, I., {et~al.} 2025, \apj,
  979, 138, \dodoi{10.3847/1538-4357/ad9920}

\bibitem[{{Harikane} {et~al.}(2023){Harikane}, {Zhang}, {Nakajima}, {Ouchi},
  {Isobe}, {Ono}, {Hatano}, {Xu}, \& {Umeda}}]{Harikane_2023_agn}
{Harikane}, Y., {Zhang}, Y., {Nakajima}, K., {et~al.} 2023, \apj, 959, 39,
  \dodoi{10.3847/1538-4357/ad029e}

\bibitem[{{Hu} {et~al.}(2025){Hu}, {Inayoshi}, {Haiman}, {Ho}, \&
  {Ohsuga}}]{Hu_2025}
{Hu}, H., {Inayoshi}, K., {Haiman}, Z., {Ho}, L.~C., \& {Ohsuga}, K. 2025,
  \apjl, 983, L37, \dodoi{10.3847/2041-8213/adc680}

\bibitem[{{Hu} {et~al.}(2022){Hu}, {Inayoshi}, {Haiman}, {Li}, {Quataert}, \&
  {Kuiper}}]{Hu_2022b}
{Hu}, H., {Inayoshi}, K., {Haiman}, Z., {et~al.} 2022, \apj, 935, 140,
  \dodoi{10.3847/1538-4357/ac7daa}

\bibitem[{{Inayoshi} \& {Ichikawa}(2024)}]{Inayoshi_Ichikawa_2024}
{Inayoshi}, K., \& {Ichikawa}, K. 2024, \apjl, 973, L49,
  \dodoi{10.3847/2041-8213/ad74e2}

\bibitem[{{Inayoshi} {et~al.}(2024){Inayoshi}, {Kimura}, \&
  {Noda}}]{Inayoshi_2025}
{Inayoshi}, K., {Kimura}, S., \& {Noda}, H. 2024, arXiv e-prints,
  arXiv:2412.03653, \dodoi{10.48550/arXiv.2412.03653}

\bibitem[{{Inayoshi} {et~al.}(2022){Inayoshi}, {Nakatani}, {Toyouchi},
  {Hosokawa}, {Kuiper}, \& {Onoue}}]{Inayoshi_2022a}
{Inayoshi}, K., {Nakatani}, R., {Toyouchi}, D., {et~al.} 2022, \apj, 927, 237,
  \dodoi{10.3847/1538-4357/ac4751}

\bibitem[{{Inayoshi} {et~al.}(2025){Inayoshi}, {Shangguan}, {Chen}, {Ho}, \&
  {Haiman}}]{Inayoshi_2025b}
{Inayoshi}, K., {Shangguan}, J., {Chen}, X., {Ho}, L.~C., \& {Haiman}, Z. 2025,
  arXiv e-prints, arXiv:2505.05322, \dodoi{10.48550/arXiv.2505.05322}

\bibitem[{{Inayoshi} {et~al.}(2020){Inayoshi}, {Visbal}, \&
  {Haiman}}]{Inayoshi_ARAA_2020}
{Inayoshi}, K., {Visbal}, E., \& {Haiman}, Z. 2020, \araa, 58, 27,
  \dodoi{10.1146/annurev-astro-120419-014455}

\bibitem[{{Ioka} \& {Nakamura}(2002)}]{Ioka_Nakamura_2002}
{Ioka}, K., \& {Nakamura}, T. 2002, \apjl, 570, L21, \dodoi{10.1086/340815}

\bibitem[{{Ji} {et~al.}(2025){Ji}, {Maiolino}, {{\"U}bler}, {Scholtz},
  {D'Eugenio}, {Sun}, {Perna}, {Turner}, {Arribas}, {Bennett}, {Bunker},
  {Carniani}, {Charlot}, {Cresci}, {Curti}, {Egami}, {Fabian}, {Inayoshi},
  {Isobe}, {Jones}, {Juod{\v{z}}balis}, {Kumari}, {Lyu}, {Mazzolari},
  {Parlanti}, {Robertson}, {Rodr{\'\i}guez Del Pino}, {Schneider}, {Sijacki},
  {Tacchella}, {Trinca}, {Valiante}, {Venturi}, {Volonteri}, {Willott},
  {Witten}, \& {Witstok}}]{Ji_2025}
{Ji}, X., {Maiolino}, R., {{\"U}bler}, H., {et~al.} 2025, arXiv e-prints,
  arXiv:2501.13082, \dodoi{10.48550/arXiv.2501.13082}

\bibitem[{{Jiang} {et~al.}(2025){Jiang}, {Jia}, {Zheng}, {Ho}, {Inayoshi},
  {Shen}, {Vogelsberger}, \& {Feng}}]{Jiang_2025}
{Jiang}, F., {Jia}, Z., {Zheng}, H., {et~al.} 2025, arXiv e-prints,
  arXiv:2503.23710, \dodoi{10.48550/arXiv.2503.23710}

\bibitem[{{Juod{\v{z}}balis} {et~al.}(2024){Juod{\v{z}}balis}, {Ji},
  {Maiolino}, {D'Eugenio}, {Scholtz}, {Risaliti}, {Fabian}, {Mazzolari},
  {Gilli}, {Prandoni}, {Arribas}, {Bunker}, {Carniani}, {Charlot},
  {Curtis-Lake}, {de Graaff}, {Hainline}, {Parlanti}, {Perna},
  {P{\'e}rez-Gonz{\'a}lez}, {Robertson}, {Tacchella}, {{\"U}bler}, {Williams},
  {Willott}, \& {Witstok}}]{Juodzbalis_2024}
{Juod{\v{z}}balis}, I., {Ji}, X., {Maiolino}, R., {et~al.} 2024, \mnras, 535,
  853, \dodoi{10.1093/mnras/stae2367}

\bibitem[{{Khan} {et~al.}(2025){Khan}, {Davis}, {Macci{\`o}}, \&
  {Holley-Bockelmann}}]{Khan_2025}
{Khan}, F.~M., {Davis}, B.~L., {Macci{\`o}}, A.~V., \& {Holley-Bockelmann}, K.
  2025, \apjl, 986, L1, \dodoi{10.3847/2041-8213/adda4c}

\bibitem[{{Kirsten} {et~al.}(2024){Kirsten}, {Ould-Boukattine}, {Herrmann},
  {Gawro{\'n}ski}, {Hessels}, {Lu}, {Snelders}, {Chawla}, {Yang}, {Blaauw},
  {Nimmo}, {Puchalska}, {Wolak}, \& {van Ruiten}}]{Kirsten_2024}
{Kirsten}, F., {Ould-Boukattine}, O.~S., {Herrmann}, W., {et~al.} 2024, Nature
  Astronomy, 8, 337, \dodoi{10.1038/s41550-023-02153-z}

\bibitem[{{Kocevski} {et~al.}(2023){Kocevski}, {Onoue}, {Inayoshi}, {Trump},
  {Arrabal Haro}, {Grazian}, {Dickinson}, {Finkelstein}, {Kartaltepe},
  {Hirschmann}, {Aird}, {Holwerda}, {Fujimoto}, {Juneau}, {Amor{\'\i}n},
  {Backhaus}, {Bagley}, {Barro}, {Bell}, {Bisigello}, {Calabr{\`o}}, {Cleri},
  {Cooper}, {Ding}, {Grogin}, {Ho}, {Hutchison}, {Inoue}, {Jiang}, {Jones},
  {Koekemoer}, {Li}, {Li}, {McGrath}, {Molina}, {Papovich},
  {P{\'e}rez-Gonz{\'a}lez}, {Pirzkal}, {Wilkins}, {Yang}, \&
  {Yung}}]{Kocevski_2023}
{Kocevski}, D.~D., {Onoue}, M., {Inayoshi}, K., {et~al.} 2023, \apjl, 954, L4,
  \dodoi{10.3847/2041-8213/ace5a0}

\bibitem[{{Kocevski} {et~al.}(2025){Kocevski}, {Finkelstein}, {Barro},
  {Taylor}, {Calabr{\`o}}, {Laloux}, {Buchner}, {Trump}, {Leung}, {Yang},
  {Dickinson}, {P{\'e}rez-Gonz{\'a}lez}, {Pacucci}, {Inayoshi}, {Somerville},
  {McGrath}, {Akins}, {Bagley}, {Bowler}, {Bisigello}, {Carnall}, {Casey},
  {Cheng}, {Cleri}, {Costantin}, {Cullen}, {Davis}, {Donnan}, {Dunlop},
  {Ellis}, {Ferguson}, {Fujimoto}, {Fontana}, {Giavalisco}, {Grazian},
  {Grogin}, {Hathi}, {Hirschmann}, {Huertas-Company}, {Holwerda},
  {Illingworth}, {Juneau}, {Kartaltepe}, {Koekemoer}, {Li}, {Lucas}, {Magee},
  {Mason}, {McLeod}, {McLure}, {Napolitano}, {Papovich}, {Pirzkal},
  {Rodighiero}, {Santini}, {Wilkins}, \& {Yung}}]{Kocevski_2025}
{Kocevski}, D.~D., {Finkelstein}, S.~L., {Barro}, G., {et~al.} 2025, \apj, 986,
  126, \dodoi{10.3847/1538-4357/adbc7d}

\bibitem[{{Kokorev} {et~al.}(2024{\natexlab{a}}){Kokorev}, {Caputi}, {Greene},
  {Dayal}, {Trebitsch}, {Cutler}, {Fujimoto}, {Labb{\'e}}, {Miller}, {Iani},
  {Navarro-Carrera}, \& {Rinaldi}}]{Kokorev_2024a}
{Kokorev}, V., {Caputi}, K.~I., {Greene}, J.~E., {et~al.} 2024{\natexlab{a}},
  \apj, 968, 38, \dodoi{10.3847/1538-4357/ad4265}

\bibitem[{{Kokorev} {et~al.}(2024{\natexlab{b}}){Kokorev}, {Chisholm},
  {Endsley}, {Finkelstein}, {Greene}, {Akins}, {Bromm}, {Casey}, {Fujimoto},
  {Labb{\'e}}, \& {Larson}}]{Kokorev_2024b}
{Kokorev}, V., {Chisholm}, J., {Endsley}, R., {et~al.} 2024{\natexlab{b}},
  \apj, 975, 178, \dodoi{10.3847/1538-4357/ad7d03}

\bibitem[{{Kormendy} \& {Ho}(2013)}]{Kormendy_Ho_2013}
{Kormendy}, J., \& {Ho}, L.~C. 2013, \araa, 51, 511,
  \dodoi{10.1146/annurev-astro-082708-101811}

\bibitem[{{Kroupa} {et~al.}(2020){Kroupa}, {Subr}, {Jerabkova}, \&
  {Wang}}]{Kroupa_2020}
{Kroupa}, P., {Subr}, L., {Jerabkova}, T., \& {Wang}, L. 2020, \mnras, 498,
  5652, \dodoi{10.1093/mnras/staa2276}

\bibitem[{{Labbe} {et~al.}(2025){Labbe}, {Greene}, {Bezanson}, {Fujimoto},
  {Furtak}, {Goulding}, {Matthee}, {Naidu}, {Oesch}, {Atek}, {Brammer},
  {Chemerynska}, {Coe}, {Cutler}, {Dayal}, {Feldmann}, {Franx}, {Glazebrook},
  {Leja}, {Maseda}, {Marchesini}, {Nanayakkara}, {Nelson}, {Pan}, {Papovich},
  {Price}, {Suess}, {Wang}, {Weaver}, {Whitaker}, {Williams}, \&
  {Zitrin}}]{Labbe_2025}
{Labbe}, I., {Greene}, J.~E., {Bezanson}, R., {et~al.} 2025, \apj, 978, 92,
  \dodoi{10.3847/1538-4357/ad3551}

\bibitem[{{Li} \& {Fenimore}(1996)}]{Li_Fenimore_1996}
{Li}, H., \& {Fenimore}, E.~E. 1996, \apjl, 469, L115, \dodoi{10.1086/310275}

\bibitem[{{Li} {et~al.}(2023){Li}, {Inayoshi}, {Onoue}, \&
  {Toyouchi}}]{Li_2023}
{Li}, W., {Inayoshi}, K., {Onoue}, M., \& {Toyouchi}, D. 2023, \apj, 950, 85,
  \dodoi{10.3847/1538-4357/accbbe}

\bibitem[{{Li} {et~al.}(2021){Li}, {Inayoshi}, \& {Qiu}}]{Li_2021}
{Li}, W., {Inayoshi}, K., \& {Qiu}, Y. 2021, \apj, 917, 60,
  \dodoi{10.3847/1538-4357/ac0adc}

\bibitem[{{Loeb} \& {Rasio}(1994)}]{Loeb_Rasio_1994}
{Loeb}, A., \& {Rasio}, F.~A. 1994, \apj, 432, 52, \dodoi{10.1086/174548}

\bibitem[{{Ma} {et~al.}(2025){Ma}, {Greene}, {Setton}, {Goulding},
  {Annunziatella}, {Fan}, {Kokorev}, {Labbe}, {Li}, {Lin}, {Marchesini},
  {Matthee}, {Robbins}, {Sajina}, {Sawicki}, \& {Telford}}]{Ma_2025}
{Ma}, Y., {Greene}, J.~E., {Setton}, D.~J., {et~al.} 2025, arXiv e-prints,
  arXiv:2504.08032, \dodoi{10.48550/arXiv.2504.08032}

\bibitem[{{Madau} \& {Haardt}(2024)}]{Madau_Haardt_2024}
{Madau}, P., \& {Haardt}, F. 2024, \apjl, 976, L24,
  \dodoi{10.3847/2041-8213/ad90e1}

\bibitem[{{Maiolino} {et~al.}(2024){Maiolino}, {Scholtz}, {Curtis-Lake},
  {Carniani}, {Baker}, {de Graaff}, {Tacchella}, {{\"U}bler}, {D'Eugenio},
  {Witstok}, {Curti}, {Arribas}, {Bunker}, {Charlot}, {Chevallard},
  {Eisenstein}, {Egami}, {Ji}, {Jones}, {Lyu}, {Rawle}, {Robertson},
  {Rujopakarn}, {Perna}, {Sun}, {Venturi}, {Williams}, \&
  {Willott}}]{Maiolino_2024_JADES}
{Maiolino}, R., {Scholtz}, J., {Curtis-Lake}, E., {et~al.} 2024, \aap, 691,
  A145, \dodoi{10.1051/0004-6361/202347640}

\bibitem[{{Maiolino} {et~al.}(2025){Maiolino}, {Risaliti}, {Signorini},
  {Trefoloni}, {Juod{\v{z}}balis}, {Scholtz}, {{\"U}bler}, {D'Eugenio},
  {Carniani}, {Fabian}, {Ji}, {Mazzolari}, {Bertola}, {Brusa}, {Bunker},
  {Charlot}, {Comastri}, {Cresci}, {DeCoursey}, {Egami}, {Fiore}, {Gilli},
  {Perna}, {Tacchella}, \& {Venturi}}]{Maiolino_2025}
{Maiolino}, R., {Risaliti}, G., {Signorini}, M., {et~al.} 2025, \mnras, 538,
  1921, \dodoi{10.1093/mnras/staf359}

\bibitem[{{Matthee} {et~al.}(2024){Matthee}, {Naidu}, {Brammer}, {Chisholm},
  {Eilers}, {Goulding}, {Greene}, {Kashino}, {Labbe}, {Lilly}, {Mackenzie},
  {Oesch}, {Weibel}, {Wuyts}, {Xiao}, {Bordoloi}, {Bouwens}, {van Dokkum},
  {Illingworth}, {Kramarenko}, {Maseda}, {Mason}, {Meyer}, {Nelson}, {Reddy},
  {Shivaei}, {Simcoe}, \& {Yue}}]{Matthee_2024}
{Matthee}, J., {Naidu}, R.~P., {Brammer}, G., {et~al.} 2024, \apj, 963, 129,
  \dodoi{10.3847/1538-4357/ad2345}

\bibitem[{{Onoue} {et~al.}(2023){Onoue}, {Inayoshi}, {Ding}, {Li}, {Li},
  {Molina}, {Inoue}, {Jiang}, \& {Ho}}]{Onoue_2023}
{Onoue}, M., {Inayoshi}, K., {Ding}, X., {et~al.} 2023, \apjl, 942, L17,
  \dodoi{10.3847/2041-8213/aca9d3}

\bibitem[{{Onoue} {et~al.}(2024){Onoue}, {Ding}, {Silverman}, {Matsuoka},
  {Izumi}, {Strauss}, {Ward}, {Phillips}, {Andika}, {Aoki}, {Arita}, {Baba},
  {Bieri}, {Bosman}, {Eilers}, {Fujimoto}, {Habouzit}, {Haiman}, {Imanishi},
  {Inayoshi}, {Ito}, {Iwasawa}, {Jahnke}, {Kashikawa}, {Kawaguchi}, {Kohno},
  {Lee}, {Li}, {Lupi}, {Lyu}, {Nagao}, {Overzier}, {Schindler}, {Schramm},
  {Scoggins}, {Shimasaku}, {Toba}, {Trakhtenbrot}, {Trebitsch}, {Treu},
  {Umehata}, {Venemans}, {Vestergaard}, {Volonteri}, {Walter}, {Wang}, {Yang},
  \& {Zhang}}]{Onoue_2024}
{Onoue}, M., {Ding}, X., {Silverman}, J.~D., {et~al.} 2024, arXiv e-prints,
  arXiv:2409.07113, \dodoi{10.48550/arXiv.2409.07113}

\bibitem[{{Oser} {et~al.}(2010){Oser}, {Ostriker}, {Naab}, {Johansson}, \&
  {Burkert}}]{Oser_2010}
{Oser}, L., {Ostriker}, J.~P., {Naab}, T., {Johansson}, P.~H., \& {Burkert}, A.
  2010, \apj, 725, 2312, \dodoi{10.1088/0004-637X/725/2/2312}

\bibitem[{{P{\'e}rez-Gonz{\'a}lez} {et~al.}(2024){P{\'e}rez-Gonz{\'a}lez},
  {Barro}, {Rieke}, {Lyu}, {Rieke}, {Alberts}, {Williams}, {Hainline}, {Sun},
  {Pusk{\'a}s}, {Annunziatella}, {Baker}, {Bunker}, {Egami}, {Ji}, {Johnson},
  {Robertson}, {Rodr{\'\i}guez Del Pino}, {Rujopakarn}, {Shivaei}, {Tacchella},
  {Willmer}, \& {Willott}}]{Perez-Gonzalez_2024}
{P{\'e}rez-Gonz{\'a}lez}, P.~G., {Barro}, G., {Rieke}, G.~H., {et~al.} 2024,
  \apj, 968, 4, \dodoi{10.3847/1538-4357/ad38bb}

\bibitem[{{Planck Collaboration} {et~al.}(2016){Planck Collaboration}, {Ade},
  {Aghanim}, {Arnaud}, {Ashdown}, {Aumont}, {Baccigalupi}, {Banday},
  {Barreiro}, {Bartlett}, {Bartolo}, {Battaner}, {Battye}, {Benabed},
  {Beno{\^\i}t}, {Benoit-L{\'e}vy}, {Bernard}, {Bersanelli}, {Bielewicz},
  {Bock}, {Bonaldi}, {Bonavera}, {Bond}, {Borrill}, {Bouchet}, {Boulanger},
  {Bucher}, {Burigana}, {Butler}, {Calabrese}, {Cardoso}, {Catalano},
  {Challinor}, {Chamballu}, {Chary}, {Chiang}, {Chluba}, {Christensen},
  {Church}, {Clements}, {Colombi}, {Colombo}, {Combet}, {Coulais}, {Crill},
  {Curto}, {Cuttaia}, {Danese}, {Davies}, {Davis}, {de Bernardis}, {de Rosa},
  {de Zotti}, {Delabrouille}, {D{\'e}sert}, {Di Valentino}, {Dickinson},
  {Diego}, {Dolag}, {Dole}, {Donzelli}, {Dor{\'e}}, {Douspis}, {Ducout},
  {Dunkley}, {Dupac}, {Efstathiou}, {Elsner}, {En{\ss}lin}, {Eriksen},
  {Farhang}, {Fergusson}, {Finelli}, {Forni}, {Frailis}, {Fraisse},
  {Franceschi}, {Frejsel}, {Galeotta}, {Galli}, {Ganga}, {Gauthier}, {Gerbino},
  {Ghosh}, {Giard}, {Giraud-H{\'e}raud}, {Giusarma}, {Gjerl{\o}w},
  {Gonz{\'a}lez-Nuevo}, {G{\'o}rski}, {Gratton}, {Gregorio}, {Gruppuso},
  {Gudmundsson}, {Hamann}, {Hansen}, {Hanson}, {Harrison}, {Helou},
  {Henrot-Versill{\'e}}, {Hern{\'a}ndez-Monteagudo}, {Herranz}, {Hildebrandt},
  {Hivon}, {Hobson}, {Holmes}, {Hornstrup}, {Hovest}, {Huang}, {Huffenberger},
  {Hurier}, {Jaffe}, {Jaffe}, {Jones}, {Juvela}, {Keih{\"a}nen}, {Keskitalo},
  {Kisner}, {Kneissl}, {Knoche}, {Knox}, {Kunz}, {Kurki-Suonio}, {Lagache},
  {L{\"a}hteenm{\"a}ki}, {Lamarre}, {Lasenby}, {Lattanzi}, {Lawrence}, {Leahy},
  {Leonardi}, {Lesgourgues}, {Levrier}, {Lewis}, {Liguori}, {Lilje},
  {Linden-V{\o}rnle}, {L{\'o}pez-Caniego}, {Lubin}, {Mac{\'\i}as-P{\'e}rez},
  {Maggio}, {Maino}, {Mandolesi}, {Mangilli}, {Marchini}, {Maris}, {Martin},
  {Martinelli}, {Mart{\'\i}nez-Gonz{\'a}lez}, {Masi}, {Matarrese}, {McGehee},
  {Meinhold}, {Melchiorri}, {Melin}, {Mendes}, {Mennella}, {Migliaccio},
  {Millea}, {Mitra}, {Miville-Desch{\^e}nes}, {Moneti}, {Montier}, {Morgante},
  {Mortlock}, {Moss}, {Munshi}, {Murphy}, {Naselsky}, {Nati}, {Natoli},
  {Netterfield}, {N{\o}rgaard-Nielsen}, {Noviello}, {Novikov}, {Novikov},
  {Oxborrow}, {Paci}, {Pagano}, {Pajot}, {Paladini}, {Paoletti}, {Partridge},
  {Pasian}, {Patanchon}, {Pearson}, {Perdereau}, {Perotto}, {Perrotta},
  {Pettorino}, {Piacentini}, {Piat}, {Pierpaoli}, {Pietrobon}, {Plaszczynski},
  {Pointecouteau}, {Polenta}, {Popa}, {Pratt}, \& {Pr{\'e}zeau}}]{Planck_2016}
{Planck Collaboration}, {Ade}, P.~A.~R., {Aghanim}, N., {et~al.} 2016, \aap,
  594, A13, \dodoi{10.1051/0004-6361/201525830}

\bibitem[{{Pouliasis} {et~al.}(2024){Pouliasis}, {Ruiz}, {Georgantopoulos},
  {Vito}, {Gilli}, {Vignali}, {Ueda}, {Koulouridis}, {Akiyama}, {Marchesi},
  {Laloux}, {Nagao}, {Paltani}, {Pierre}, {Toba}, {Habouzit}, {Vijarnwannaluk},
  \& {Garrel}}]{Pouliasis_2024}
{Pouliasis}, E., {Ruiz}, A., {Georgantopoulos}, I., {et~al.} 2024, \aap, 685,
  A97, \dodoi{10.1051/0004-6361/202348479}

\bibitem[{{Reines} \& {Volonteri}(2015)}]{Reines_Volonteri_2015}
{Reines}, A.~E., \& {Volonteri}, M. 2015, \apj, 813, 82,
  \dodoi{10.1088/0004-637X/813/2/82}

\bibitem[{{Scoggins} \& {Haiman}(2024)}]{Scoggins_2024}
{Scoggins}, M.~T., \& {Haiman}, Z. 2024, \mnras, 531, 4584,
  \dodoi{10.1093/mnras/stae1449}

\bibitem[{{Shang} {et~al.}(2010){Shang}, {Bryan}, \& {Haiman}}]{Shang_2010}
{Shang}, C., {Bryan}, G.~L., \& {Haiman}, Z. 2010, \mnras, 402, 1249,
  \dodoi{10.1111/j.1365-2966.2009.15960.x}

\bibitem[{{Tanaka} {et~al.}(2024){Tanaka}, {Silverman}, {Shimasaku}, {Arita},
  {Akins}, {Inayoshi}, {Ding}, {Onoue}, {Liu}, {Casey}, {Lambrides}, {Kokorev},
  {Jin}, {Faisst}, {Drakos}, {Shen}, {Li}, {Zhuang}, {Fei}, {Ito}, {Ren},
  {Matsui}, {Ando}, {Hatano}, {Fujii}, {Kartaltepe}, {Koekemoer}, {Liu},
  {McCracken}, {Rhodes}, {Robertson}, {Franco}, {Andika}, {Cloonan}, {Fan},
  {Gozaliasl}, {Harish}, {Hayward}, {Huertas-Company}, {Kakkad}, {Kinugawa},
  {Roy}, {Shuntov}, {Talia}, {Toft}, {Vijayan}, \& {Zhang}}]{Tanaka_2024b}
{Tanaka}, T.~S., {Silverman}, J.~D., {Shimasaku}, K., {et~al.} 2024, arXiv
  e-prints, arXiv:2412.14246, \dodoi{10.48550/arXiv.2412.14246}

\bibitem[{{Taylor} {et~al.}(2025{\natexlab{a}}){Taylor}, {Finkelstein},
  {Kocevski}, {Jeon}, {Bromm}, {Amor{\'\i}n}, {Arrabal Haro}, {Backhaus},
  {Bagley}, {Banados}, {Bhatawdekar}, {Brooks}, {Calabr{\`o}}, {Ortiz},
  {Cheng}, {Cleri}, {Cole}, {Davis}, {Dickinson}, {Donnan}, {Dunlop}, {Ellis},
  {Fern{\'a}ndez}, {Fontana}, {Fujimoto}, {Giavalisco}, {Grazian}, {Guo},
  {Hathi}, {Holwerda}, {Hirschmann}, {Inayoshi}, {Kartaltepe}, {Khusanova},
  {Koekemoer}, {Kokorev}, {Larson}, {Leung}, {Lucas}, {McLeod}, {Napolitano},
  {Onoue}, {Pacucci}, {Papovich}, {P{\'e}rez-Gonz{\'a}lez}, {Pirzkal},
  {Somerville}, {Trump}, {Wilkins}, {Yung}, \& {Zhang}}]{Taylor_2025a}
{Taylor}, A.~J., {Finkelstein}, S.~L., {Kocevski}, D.~D., {et~al.}
  2025{\natexlab{a}}, \apj, 986, 165, \dodoi{10.3847/1538-4357/add15b}

\bibitem[{{Taylor} {et~al.}(2025{\natexlab{b}}){Taylor}, {Kokorev}, {Kocevski},
  {Akins}, {Cullen}, {Dickinson}, {Finkelstein}, {Arrabal Haro}, {Bromm},
  {Giavalisco}, {Inayoshi}, {Juneau}, {Leung}, {Perez-Gonzalez}, {Somerville},
  {Trump}, {Amorin}, {Barro}, {Burgarella}, {Brooks}, {Carnall}, {Casey},
  {Cheng}, {Chisholm}, {Chworowsky}, {Davis}, {Donnan}, {Dunlop}, {Ellis},
  {Fernandez}, {Fujimoto}, {Grogin}, {Gupta}, {Hathi}, {Jung}, {Hirschmann},
  {Kartaltepe}, {Koekemoer}, {Larson}, {Leung}, {Llerena}, {Lucas}, {McLeod},
  {McLure}, {Napolitano}, {Papovich}, {Stanton}, {Tripodi}, {Wang}, {Wilkins},
  {Yung}, \& {Zavala}}]{Taylor_2025b}
{Taylor}, A.~J., {Kokorev}, V., {Kocevski}, D.~D., {et~al.} 2025{\natexlab{b}},
  arXiv e-prints, arXiv:2505.04609, \dodoi{10.48550/arXiv.2505.04609}

\bibitem[{{Thomas} {et~al.}(2005){Thomas}, {Maraston}, {Bender}, \& {Mendes de
  Oliveira}}]{Thomas_2005}
{Thomas}, D., {Maraston}, C., {Bender}, R., \& {Mendes de Oliveira}, C. 2005,
  \apj, 621, 673, \dodoi{10.1086/426932}

\bibitem[{{Trinca} {et~al.}(2022){Trinca}, {Schneider}, {Valiante}, {Graziani},
  {Zappacosta}, \& {Shankar}}]{Trinca_2022}
{Trinca}, A., {Schneider}, R., {Valiante}, R., {et~al.} 2022, \mnras, 511, 616,
  \dodoi{10.1093/mnras/stac062}

\bibitem[{{Trujillo} {et~al.}(2009){Trujillo}, {Cenarro}, {de
  Lorenzo-C{\'a}ceres}, {Vazdekis}, {de la Rosa}, \& {Cava}}]{Trujillo_2009}
{Trujillo}, I., {Cenarro}, A.~J., {de Lorenzo-C{\'a}ceres}, A., {et~al.} 2009,
  \apjl, 692, L118, \dodoi{10.1088/0004-637X/692/2/L118}

\bibitem[{{Ueda} {et~al.}(2014){Ueda}, {Akiyama}, {Hasinger}, {Miyaji}, \&
  {Watson}}]{Ueda_2014}
{Ueda}, Y., {Akiyama}, M., {Hasinger}, G., {Miyaji}, T., \& {Watson}, M.~G.
  2014, \apj, 786, 104, \dodoi{10.1088/0004-637X/786/2/104}

\bibitem[{{Valiante} {et~al.}(2016){Valiante}, {Schneider}, {Volonteri}, \&
  {Omukai}}]{Valiante_2016}
{Valiante}, R., {Schneider}, R., {Volonteri}, M., \& {Omukai}, K. 2016, \mnras,
  457, 3356, \dodoi{10.1093/mnras/stw225}

\bibitem[{{Volonteri} {et~al.}(2021){Volonteri}, {Habouzit}, \&
  {Colpi}}]{Volonteri_2021}
{Volonteri}, M., {Habouzit}, M., \& {Colpi}, M. 2021, Nature Reviews Physics,
  3, 732, \dodoi{10.1038/s42254-021-00364-9}

\bibitem[{{Wang} {et~al.}(2024){Wang}, {Leja}, {de Graaff}, {Brammer},
  {Weibel}, {van Dokkum}, {Baggen}, {Suess}, {Greene}, {Bezanson}, {Cleri},
  {Hirschmann}, {Labb{\'e}}, {Matthee}, {McConachie}, {Naidu}, {Nelson},
  {Oesch}, {Setton}, \& {Williams}}]{Wang_2024b}
{Wang}, B., {Leja}, J., {de Graaff}, A., {et~al.} 2024, \apjl, 969, L13,
  \dodoi{10.3847/2041-8213/ad55f7}

\bibitem[{{Wang} {et~al.}(2025){Wang}, {de Graaff}, {Davies}, {Greene}, {Leja},
  {Brammer}, {Goulding}, {Miller}, {Suess}, {Weibel}, {Williams}, {Bezanson},
  {Boogaard}, {Cleri}, {Hirschmann}, {Katz}, {Labb{\'e}}, {Maseda}, {Matthee},
  {McConachie}, {Naidu}, {Oesch}, {Rix}, {Setton}, \& {Whitaker}}]{Wang_2025}
{Wang}, B., {de Graaff}, A., {Davies}, R.~L., {et~al.} 2025, \apj, 984, 121,
  \dodoi{10.3847/1538-4357/adc1ca}

\end{thebibliography}


\end{document}